\documentclass[english,prl,twocolumn,showpacs,preprintnumbers,amsmath,amssymb,a4paper]{revtex4}
\usepackage[T1]{fontenc}
\usepackage[latin1]{inputenc}
\usepackage{graphicx}
\usepackage[top=0.8in,bottom=0.9in,,left=0.7in,right=0.7in]{geometry}
\usepackage{sidecap}

\makeatletter
\usepackage{dcolumn}
\usepackage{bm}
\usepackage{epsfig}

\newcommand{\ignore}[1]{}

\usepackage{babel}
\makeatother

\begin{document}

\title{Quantum Anomalous Hall Effect in 2D Organic Topological Insulators}

\author{Z. F. Wang}
\affiliation{Department of Materials Science and Engineering,
University of Utah, Salt Lake City, UT 84112, USA}

\author{Zheng Liu}
\affiliation{Department of Materials Science and Engineering,
University of Utah, Salt Lake City, UT 84112, USA}

\author{Feng Liu}
\thanks{Corresponding author. E-mail: fliu@eng.utah.edu}
\affiliation{Department of Materials Science and Engineering,
University of Utah, Salt Lake City, UT 84112, USA}

\begin{abstract}
Quantum anomalous Hall effect (QAHE) is a fundamental transport phenomenon in
the field of condensed-matter physics. Without external magnetic field,
spontaneous magnetization combined with spin-orbit coupling give rise
to a quantized Hall conductivity. So far, a number of theoretical proposals
have been made to realize the QAHE, but all based on {\it inorganic} materials.
Here, using first-principles calculations, we predict a family of 2D {\it organic}
topological insulators for realizing the QAHE. Designed by assembling
molecular building blocks of triphenyl-transition-metal compounds into a
hexagonal lattice, this new class of organic materials are shown to have
a nonzero Chern number and exhibit a gapless chiral edge state within the Dirac gap.
\end{abstract}


\maketitle
The quantum Hall effect refers to the quantized Hall conductivity due to Landau quantization,
as observed in a 2D electron system \cite{2}. The essential ingredient to produce
the quantum Hall effect is to break the time-reversal symmetry, usually by applying
an external magnetic field.
An interesting alternative is to have the internal magnetization coupled with
spin-orbit coupling (SOC) that can also break time-reversal symmetry without magnetic field \cite{3}.
This is called quantum anomalous Hall effect (QAHE), as first proposed by Haldane \cite{5}.
Subsequently, some realistic
materials were theoretically proposed to realize the QAHE, such as mercury-based
quantum wells \cite{6}, graphene \cite{7,8,9} and topological insulators (TIs) \cite{10}.
In general, in a QAHE material, internal magnetism, such as the one induced by transition
metals, breaks the time-reversal symmetry and splits the spin degenerated bands.
In addition, the SOC opens a global band gap, giving rise to a topologically nontrivial
insulating state characterized with the quantized Hall conductivity.

Due to the difficulty in controlling magnetization and SOC,
the QAHE has yet to be observed experimentally \cite{11,12}. So far, all the theoretical proposals
for realizing the QAHE are based on inorganic materials. It is fascinating to note that many
fundamental physical phenomena in inorganic materials and devices have always found their way
to organic counterparts, such as the organic superconductors \cite{13},
light-emitting diodes \cite{14}, solar cells \cite{15} and field-effect transistors \cite{16}.
Therefore, an interesting question is whether the QAHE can be realized in organic materials.

\begin{figure}[htpb]
\begin{center}
\epsfig{figure=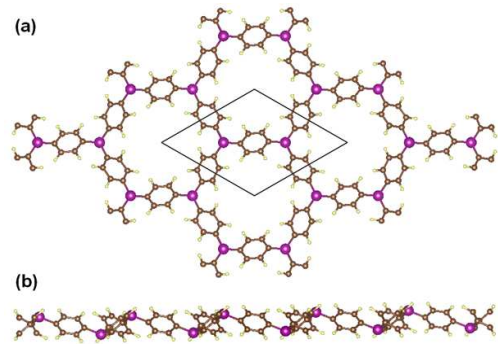,width=6.5cm}
\end{center}
\caption{(a) Top and (b) side
view of the optimized 2D TMn lattice structure. Rhombus shows the unit cell.}
\label{fig:fig-1}
\end{figure}

In this Letter, we demonstrate that QAHE can indeed be realized in 2D organic TIs (OTIs) self-assembled from
triphenyl-transition-metal compounds, using triphenyl-manganese (TMn) as a model system.
Based on Chern number and edge state calculations, we confirm the 2D TMn lattice have
nontrivial topological Dirac-gap states and explain the physical origin of its QAHE due to
both the intrinsic SOC and strong magnetization provided by Mn atoms.

Our first-principles band structure and band topology
calculations were carried out in the framework of the PBE-GGA functional using the
VASP package \cite{19}. All self-consistent calculations were performed with a plane-wave
cutoff of 500 eV on a $7\times7\times1$ Monkhorst-Pack k-point mesh. Supercell with a vacuum
layer more than 15 {\AA} thick is used to ensure decoupling between neighboring slabs.
For structural relaxation, all the atoms are allowed to relax until atomic forces
are smaller than 0.01 eV/{\AA}.

The TMn [Mn(C$_6$H$_5$)$_3$] molecule consists of a Mn atom bonded with three benzene rings with three-fold
rotational symmetry. When bridged with a Mn atom, they may naturally form a 2D hexagonal
lattice, as shown in Fig. 1. There are two Mn atoms and three benzene rings with a chemical formula
of Mn$_2$C$_{18}$H$_{12}$ in each unit cell and the neighboring benzene rings are bridge-bonded
through the para-Mn atoms. The optimized 2D lattice is buckled with the para-Mn atoms moving
alternately up and down out of the plane of benzene rings (see Fig. 1b). In addition, each
benzene rotates slightly along the Mn-Mn axis in a clockwise manner. The equilibrium lattice
constant is found to be 10.7 {\AA} with the Mn-Mn distance and height difference
being 6.45 {\AA} and 1.86 {\AA}, respectively.

\begin{figure}[htpb]
\begin{center}
\epsfig{figure=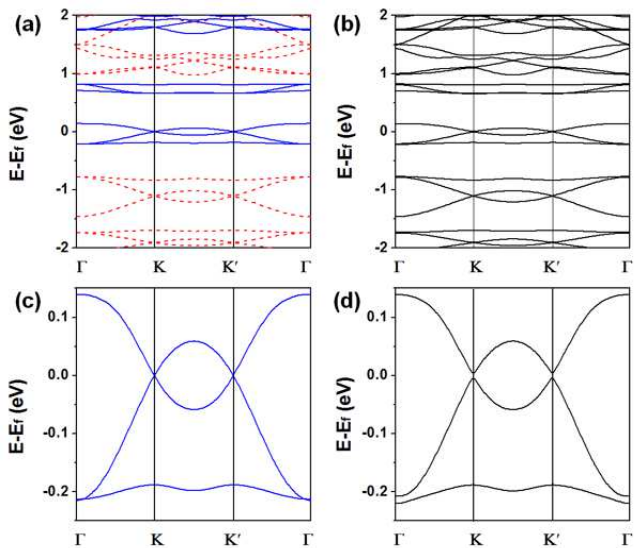,width=8.5cm}
\end{center}
\caption{(a) and (b) Band structures of the TMn lattice without and with SOC, respectively.
Red dashed lines and blue solid lines denote spin-up and spin-down bands.
(c) and (d) Magnification of (a) and (b) around
the Fermi level, respectively.}
\label{fig:fig-2}
\end{figure}

First, we analyze the band structure of TMn lattice purposely without SOC.
The ground state of TMn lattice is found ferromagnetic with a magnetic moment
of 4 $\mu_B$ per unit cell, which is 0.18 eV lower than the anti-ferromagnetic state.
The 3d-shell of Mn is half filled, after providing three d electrons to bond with the C,
each Mn atom is left with two unpaired d electrons of the same spin.
The band structure of the ferromagnetic state is shown in Fig. 2(a). Due to
internal magnetization, the spin-up (red dashed line) and spin-down (blue solid line)
bands are completely split away from each other, and only the spin-down band is left around
the Fermi level (a half semimetal). Magnifying the bands around the Fermi level, we can see a
clear linear Dirac band, with the Fermi level located exactly at the Dirac point ($K$ and $K^\prime$),
as shown in Fig. 2(c). Similar spin-up Dirac bands are below the Fermi level. Next, the SOC is included,
and the corresponding band structure is shown in Fig. 2(b). Our magnetic anisotropy calculation shown that
the SOC ground state has the out-of-plane spin orientation, which is 0.5 meV lower than the in-plane spin
orientation. Comparing Fig. 2(b) to Fig. 2(a), they are almost the same, except a sizable bulk band
gap (9.5 meV) opened at $K$ and $K^\prime$ Dirac points around Fermi level, which can be see more
clearly from the magnified band in Fig. 2(d) compared to Fig. 2(c).

\begin{figure}[htpb]
\begin{center}
\epsfig{figure=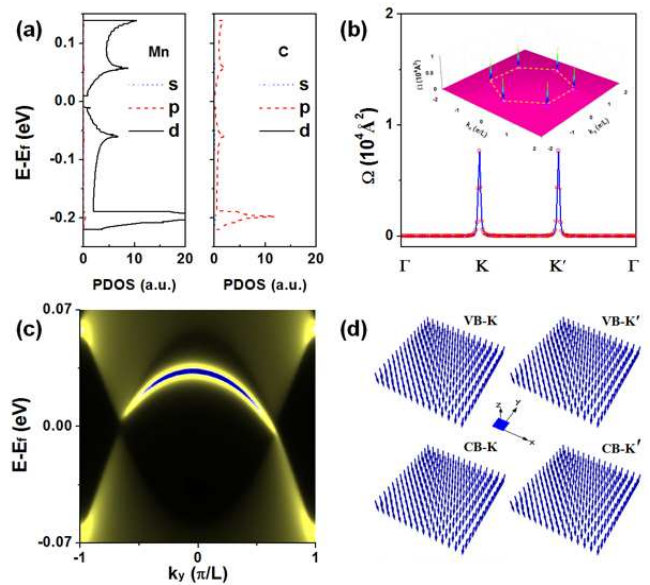,width=8.5cm}
\end{center}
\caption{
(a) Projected density of states of Dirac bands around the
Fermi level. (b) Berry curvature for the whole valence bands
(blue solid line) and the highest occupied valence band below the Fermi level (red open circle) along
the high-symmetry directions. Inset: the 2D distribution of berry curvature for the whole
valence bands in the momentum space. The first Brillouin zone is denoted by the dashed line.
(c) Semi-infinite edge states inside the Dirac gap of the TMn lattice. (d) 3D spin texture
round $K$ and $K^\prime$ for the highest occupied valence band and the lowest unoccupied conduction band.}
\label{fig:fig-3}
\end{figure}

We checked that the Dirac states mainly come from the
Mn d orbit, with little contribution from the C p orbit, as shown in Fig. 3(a). Therefore,
the Dirac band around the Fermi level originates from the
hexagonal Mn lattice. To identify the topological properties of the Dirac gap,
we calculated the Berry curvature of the bands using the Kubo formula \cite{20,21},
\begin{equation}
\begin{aligned}
&\Omega(\textbf{k})= \sum_nf_n\Omega_n(\textbf{k}), \\
\Omega_n(\textbf{k})= -\sum_{n^\prime\neq n}& 2Im
\frac {\langle\psi_{n\textbf{k}}|\upsilon_x|\psi_{n^\prime \textbf{k}}\rangle
\langle\psi_{n^\prime\textbf{k}}|\upsilon_y|\psi_{n\textbf{k}}\rangle}
{(\varepsilon_{n^\prime \textbf{k}}-\varepsilon_{n \textbf{k}})^2},
\end{aligned}
\end{equation}
where $n$ is the band index, $\varepsilon_{n \textbf{k}}$ and $\psi_{n\textbf{k}}$ are eigenvalue
and eigenstate of band $n$, respectively, $\upsilon_{x/y}$ is the velocity operator, $f_n$ is Fermi distribution
function. We have used the
maximally localized Wannier functions (MLWFs) to calculate the Berry curvature by using the
Wannier90 package \cite{22}. The inner energy window used to accurately reproduce the
first-principles bands is set from $E_f-8.0$ eV to $E_f+3.0$ eV.
Fig. 3(b) shows the Berry curvature for the whole valence
bands (blue solid line) along
the high-symmetry directions and the corresponding 2D distribution in momentum space (inset).
We see that the nonzero Berry curvatures are localized
around $K$ and $K^\prime$ points with the same sign. Integrating the Berry curvatures over the
first Brillouin zone, we obtain the Chern number, $C=\frac{1}{2\pi} \int_{BZ}d^2k\Omega=1$,
with each Dirac cone ($K$ and $K^\prime$) contributing 0.5. Such nonvanishing Chern number
characterizes a quantized Hall conductivity and confirms the QAHE in the TMn lattice.

The nonzero Chern number can also be manifested by the presence of chiral edge states within
the topological nontrivial Dirac gap. By using the MLWFs obtained from the first-principles calculation, the edge Green's
function \cite{23} of the semi-infinite TMn lattice is constructed and the local density of state (LDOS)
of edge states are calculated \cite{OTI,flat-band}, as shown
in Fig. 3(c) for one of the edges (the LDOS of the other edge state is symmetric to this one). The number of the edge states
indicates the absolute value of the Chern number. Apparently, the one chiral edge state observed in the
Dirac gap is consistent with the calculated Chern number $C=1$.

To better understand the odd Chern number $C=1$, the Berry curvature
for the highest occupied valence band (VB) below the Fermi level is also calculated, as shown
in Fig. 3(b) (red open circle), which completely matches the total Berry curvature curve. This
indicates that the sum of Chern number for other valence bands below the highest occupied valence band
is zero. Thus, we can focus on the highest occupied valence band only, and examine its
spin components around the Dirac gap. As a comparison, the lowest unoccupied conduction band (CB) above the Fermi level is
also studied. As shown in Fig. 3(d), the spin textures for VB and CB are uniformly pointing along the $-z$
direction around $K$ and $K^\prime$, as both come from the spin-down bands.
Qiao \textit{et al}. \cite{8} recently showed that such spin texture
has no contribution to the Chern number, but the pseudo-spin texture of the Dirac states may contribute to the Chern
number. To see whether the nontrivial topology of our TMn lattice is also originated from the pseudo-spin texture,
we construct a tight-binding (TB) model, as follows.

Since our TMn lattice has inversion symmetry, the gap opening at the Dirac point cannot
be induced by the Rashba SOC as in the doped graphene \cite{7,8,9}, but instead by the intrinsic SOC of Mn atoms.
To further illustrate the intrinsic SOC effect, we write a single $\pi$-band TB model
to describe the Dirac bands of hexagonal TMn lattice. To the first-order approximation, the TB
Hamiltonian with exchange field and intrinsic SOC can be written as,
\begin{equation}
\begin{split}
H=&t\sum_{\langle i,j \rangle, \alpha}c^+_{i\alpha}c_{j\alpha}-M\sum_{i,\alpha,\beta}c^+_{i\alpha}s^z_{\alpha\beta}c_{i\beta} +\\
&\lambda_{so}\frac{2i}{\sqrt{3}}\sum_{\langle\langle i,j \rangle\rangle}c^+_i\vec{s}\cdot(\vec{d}_{kj}\times\vec{d}_{ik})c_j.
\end{split}
\end{equation}
Here, $c^+_{i\alpha}$ and $c_{i\alpha}$ are $\pi$-band creation and annihilation operators, respectively, for an
electron with spin $\alpha$ on site $i$. The first term is the nearest-neighbor hopping with magnitude $t$.
The second term is the exchange field with magnitude $M$. The third term is the next-nearest-neighbor
intrinsic SOC with amplitude $\lambda_{so}$, $\vec{s}$ is the spin Pauli matrix, $\vec{d}_{kj}$ is the
unit vector pointing from site $j$ to $k$. We note that this intrinsic SOC term forbids mixing of spin-up and spin-down
states due to its special nature of 2D geometry, a condition that will breakdown if the structure is not perfect 2D.

\begin{figure}[htpb]
\begin{center}
\epsfig{figure=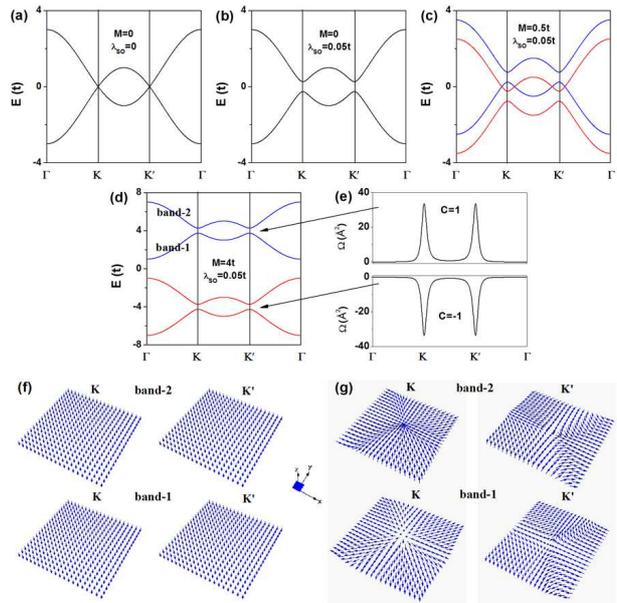,width=8.5cm}
\end{center}
\caption{TB Band structure with (a) M=0 and $\lambda_{so}$=0, (b) M=0 and $\lambda_{so}$=0.05t,
(c) M=0.5t and $\lambda_{so}$=0.05t and (d) M=4t and $\lambda_{so}$=0.05t. (e) Berry curvatures by
setting the Fermi level within the intrinsic SOC gap indicated by the arrows and the resulting Chern
numbers. Red (blue) color denotes spin-up (spin-down) bands. (f) and (g) 3D spin and
pseudo-spin texture around $K$ and $K^\prime$ for band-1 and band-2 shown in (d).}
\label{fig:fig-4}
\end{figure}

Diagonalizing the above Hamiltonian in reciprocal space, we obtain its band structure,
as shown in Fig. 4. In our calculation, all the parameters are scaled with the hopping
parameter $t$. Without SOC and exchange field, the corresponding band
structure along the high-symmetry directions is shown in Fig. 4(a), where we can see the
linear Dirac bands at $K$ and $K^\prime$ points. When we turn on the SOC but without
the exchange field, a band gap is opened at the Dirac points, as shown in Fig. 4(b). Without the
exchange field, all the bands are spin degenerated in Fig. 4(b). When we turn on both
SOC and exchange field, the spin degeneracy is lifted. If the exchange field is large enough
to overcome the SOC gap, the spin-up (red line) and spin-down (blue line) bands will
cross over with each other, as shown in Fig. 4(c). The situation here, however, is different from
the case of Rashba SOC in doped graphene \cite{7,8,9}, as there is no SOC gap opening at the band crossing points with different
spins in the physical regime of small magnetization. However, if we further increase the exchange
field, the spin-up and spin-down bands will completely separate from each other, as shown in Fig. 4(d),
which corresponds to our case of TMn lattice. Then, we have a global gap induced by intrinsic SOC,
leading to QAHE. To further check the topological properties of the gap states induced by intrinsic SOC, we calculated
the Berry curvature by setting the Fermi level inside the spin-up and spin-down Dirac gap, respectively
(The real TMn lattice has the Fermi level in the gap of spin-down bands). The corresponding Berry
curvature and Chern number are shown in Fig. 4(e). We found the non-zero Berry curvature that localizes
around the Dirac point and the $C=-1$ ($C=1$) for spin-up (spin-down) bands, consistent with our first
principles calculations. Thus, we confirm that the intrinsic SOC in TMn lattice is responsible for
gap opening to realize the QAHE.

To reveal the difference between spin and pseudo-spin, we calculate the Chern number (topological charge) resulting
from the spin and pseudo-spin texture separately, using the the TB model. The spin and pseudo-spin
components are defined as \cite{8} $\langle s_i \rangle=\langle \psi|I \otimes s_i |\psi \rangle$ and
$\langle \sigma_i \rangle=\langle \psi|\sigma_i \otimes I |\psi \rangle$, respectively, where $I$ is a 2$\times$2
identity matrix, $s_i$ ($\sigma_i$) is the spin (pseudo-spin) Pauli matrix with $i=x,y,z$ and
$|\psi \rangle$ is the eigenvector of Eq. (2) in reciprocal space. Here, we choose the spin-down bands
labeled with band-1 and band-2 in Fig. 4(d) for our calculation, as our first-principles calculations show only
spin-down bands at the Fermi Level [Fig. 2(a)].
Fig. 4(f) shows the TB spin texture with all spins uniformly pointing along the
$-z$ direction, consistent with the first-principles results [Fig. 3(d)]. However, the pseudo-spin
texture is not uniform, as shown in Fig. 4(g). Its in-plane components have different patterns
depending on the valleys ($K$ and $K^\prime$) and bands (band-1 and band-2), while its out-of-plane
components only exist near the valley and point along either $-z$ or $z$ direction. The spin
and pseudo-spin Chern number can be calculated as \cite{8,8-1}
$n=1/4\pi\int dk^2(\partial_{k_x}\hat{\textbf{h}} \times \partial_{k_y}\hat{\textbf{h}})\cdot \hat{\textbf{h}}$,
where $\hat{\textbf{h}}=\textbf{h}/|\textbf{h}|$ with $\textbf{h}$
representing the projection of Hamiltonian in Eq. (2) into spin or pseudo-spin space.
Physically, the unit vector $\hat{\textbf{h}}$ represents the expectation value of the orientation of the
spin or pseudo-spin associated with the eigenvector $|\psi \rangle$.
For band-1 in Fig. 4(d), we found $n_{spin}$=0 and $n_{pseudo}=0.5$ for both valleys.
The pseudo-spin texture provides a half topological charge, corresponding to a meron. Thus, by counting
two valleys' contribution, the Chern number $C=1$. In addition, the
sign of the topological charge is determined by the details of the pseudo-spin texture. For band-2 in Fig. 4(d),
we found $n_{spin}$=0 and $n_{pseudo}=-0.5$ for both valleys. Therefore, the total Chern number for band-1 and band-2
is zero, which is consistent with our first-principles result.

Besides realizing the QAHE in a new organic molecular lattice, it is
important to point out some new physics in our system in comparison with previous works.
The QAHE is our proposed TMn lattice has an odd Chern number ($C=1$) which is induced by intrinsic SOC
taking place in the strong magnetization
regime, while the QAHE in the transition metal doped graphene has an even Chern number ($C=2$) which
is induced by Rashba SOC taking place in the weak magnetization regime.
It was shown that the d-states in 5d transition metal doped graphene plays a
determining role in realizing the QAHE and there exists also a global Dirac gap between the bands of the same
spin [Fig. 2(c) in Ref. 7], but it is 0.27 eV below the Fermi level and its Chern number is even ($C=-2$).

Lastly, one critical point is whether the topological properties of TMn lattice can remain on a substrate.
Naturally one should look for a substrate with minimal interfacial interaction with the TMn lattice.
To test out this possibility, we have placed the TMn lattice on top of graphene [Fig. S1(a)] \cite{supporting},
which is expected to have a weak van der Waals interfacial interaction and right hexagonal symmetry. Our
calculations show that because of lattice mismatch, the TMn lattice becomes a flat structure instead of the
freestanding buckled structure, but the main features of QAHE remain intact. There is still a SOC gap at the
Dirac point around Fermi level [Figs. S1(e) and (f)], within which a nontrivial topological edge state
resides [Fig. S1(g)]. These results demonstrate the feasibility of attaining the QAHE of TMn lattice on a
substrate.

This work was supported by US DOE-BES (Grant No. DE-FG02-04ER46148) and NSF MRSEC (Grant No. DMR-1121252).
Z.F.W. acknowledges additional support from ARL (Cooperative Agreement No. W911NF-12-2-0023).
We thank the CHPC at University of Utah and NERSC for providing the computing resources.

\end{document}